\documentclass[12pt,twoside,]{article}   %

\usepackage[citestyle=numeric-comp, bibstyle=chem-rsc, articletitle=true, biblabel=brackets,doi=true,  backrefstyle=three,minnames=6,maxnames=6, sorting=none]{biblatex}
\usepackage{textcomp}
\usepackage{subfiles}

\usepackage[section]{placeins}   %
\usepackage{graphicx}
\usepackage{siunitx}
\makeatletter \renewcommand\@biblabel[1]{$^{#1}$} \makeatother
\newcommand{\cen}[1]{\begin{center} #1 \end{center}}

\usepackage{hyperref}
\hypersetup{ colorlinks,
    citecolor=blue,
    filecolor=blue,
    linkcolor=blue,
    urlcolor=blue
}
\usepackage{cleveref}

\usepackage[all]{hypcap}    

\usepackage{graphicx}
\usepackage{booktabs} 

\usepackage{comment}

\begin{document}

\cen{\sf {\Large {\bfseries Technical Note: Low-Dose Simulation for Grating-Based X-Ray Dark-Field Radiography Using a Virtually Decreased Irradiation Area} \\  
\vspace*{10mm}
Henriette Bast\textsuperscript{a,b,c}, Rafael C. Schick\textsuperscript{a,b, c}, Thomas Koehler\textsuperscript{d,e}, Franz Pfeiffer\textsuperscript{a,b,c,e}} \\
\vspace{2ex}
\textsuperscript{a} Chair of Biomedical Physics, Department of Physics, School of Natural Sciences, Technical University of Munich, 85748 Garching, Germany\linebreak
\textsuperscript{b} Munich Institute of Biomedical Engineering, 85748 Garching, Germany \linebreak
\textsuperscript{c} Institute for Diagnostic and Interventional Radiology, School of Medicine and Health, TUM Klinikum, Technical University of Munich (TUM), 81675  Munich\linebreak
\textsuperscript{d} Philips Innovative Technologies, 22335 Hamburg, Germany \linebreak
\textsuperscript{e}  Munich Institute for Advanced Study, Technical University of Munich, 85748 Garching, Germany 
}

\pagenumbering{roman}
\setcounter{page}{1}
\pagestyle{plain}
Corresponding e-mail address: henriette.bast@tum.de

Corresponding author address: Henriette Bast, Chair of Biomedical Physics, Department of Physics, School of Natural Sciences, Technical University of Munich, 85748 Garching, Germany \\

\vspace{0.2cm}

\begin{abstract} 
\noindent {\bf Background:} X-ray dark-field radiography uses small-angle scattering to visualize the structural integrity of lung alveoli.  
To study the influence of dose reduction on clinical dark-field radiographs, one can simulate low-dose images by virtually reducing the irradiated area. However, these simulations can exhibit stripe artifacts.\\
{\bf Purpose:} Validation of the low-dose simulation algorithm reported by Schick \& Bast et al., PLoS ONE, 2024 in \cite{schick_simulated_2024}. Furthermore, we want to demonstrate that stripe artifacts observed in simulated images at very low-dose levels are introduced by limitations of the algorithm and would not appear in actual low-dose dark-field images. 
\\
{\bf Methods:} Dark-field radiographs of an anthropomorphic chest phantom were acquired at different tube currents equaling different radiation doses. Based on the measurement with a high radiation dose, dark-field radiographs corresponding to lower radiation doses were simulated by virtually reducing the irradiated area. 
The simulated low-dose radiographs were evaluated by a quantitative comparison of the dark-field signal using different regions of interests.
{\bf Results:} Dark-field radiographs acquired at one quarter of the standard dose were artifact-free. The dark-field signal differed from the simulated radiographs by up to 10\%. Algorithm-induced stripe artifacts decrease the image quality of the simulated radiographs. \\
{\bf Conclusions:} Virtually reducing the irradiation area is a simple approach to generate low-dose radiographs based on images acquired with scanning-based dark-field radiography. However, as the algorithm creates stripe artifacts in the dark-field images, particularly at higher dose reductions, that are not present in measured low-dose images, simulated images have reduced image quality compared to their measured counterparts. \\

\end{abstract}

\newpage

\setlength{\baselineskip}{0.7cm}      	

\pagenumbering{arabic}
\setcounter{page}{1}
\section{Introduction}

Grating-based X-ray dark-field radiography has shown significant potential in human chest imaging  for the diagnosis of pulmonary diseases \cite{Willer2021,urban_qualitative_2022, frank_darkfield_2022, urban_darkfield_2023a}. 
The dark-field image arises from small angle scattering of X-rays as they pass through  microstructures such as pulmonary alveoli \cite{pfeiffer_hardxray_2008}, whereas the attenuation image is produced predominately by the photoelectric absorption and Compton scattering of the X-rays.
A clinical prototype system for grating-based X-ray dark-field imaging has been developed at our institution and located at the university hospital Rechts der Isar, Munich, Germany.

Since clinical dark-field radiography is a novel imaging procedure, the optimal balance between radiation dose and image quality has yet to be established.
In accordance with the ALARA principle (As Low As Reasonably Achievable),  simulations were initially performed to assess the impact of reduced radiation doses on image quality in dark-field radiographs, thereby paving the way for actual dose reduction in human dark-field imaging. Such simulated dose reduction has already been proven to enable dark-field imaging for the diagnosis of the Coronavirus SARS-CoV-2 (COVID-19) pneumonia with a\SI{75}{\%}-reduced median radiation dose of 10.5\,\textmu Sv, as reported in our earlier study \cite{schick_simulated_2024}.
One key conclusion from that work was that further dose reduction is feasible, as algorithm-related artifacts became increasingly prominent at dose reductions of \SI{75}{\%} or more, significantly reducing image quality. In this study, we build on those findings by directly comparing simulated low-dose radiographs with corresponding actual low-dose measurements, acquired using a phantom, thus validating the reported low-dose simulation algorithm.

\section{Materials and Methods}
The clinical prototype dark-field system located at the Technical University of Munich university hospital Rechts der Isar, Germany was used to to acquire attenuation and dark-field radiographs of the human thorax. The prototype's main components are three optical gratings placed between an X-ray tube (MRC 200 0508 ROT-GS 1003; Philips Medical Systems, Germany) operated at \SI{70}{kVp} and a flat-panel detector (PIXIUM 4343 F4; Trixell, France).

\subsection{Image Acquisition and Reconstruction with the Clinical Dark-field Scanning System}
The clinical prototype system is a scanning system, that covers the full field-of-view (42\,cm\texttimes 42\,cm) during image acquisition through a vertical scan of 6.5\,s duration with an illuminated grating are of 42\,cm\texttimes 6.5\,cm \cite{Willer2021, frank_darkfield_2022}. During this scanning procedure, a series of 195 exposures are measured by the detector at a frame rate of 30\,Hz. 
After image reconstruction using a phase retrieval algorithm, both the simulated reduced dose images and the actual, experimental images were corrected for different sources of scatter \cite{urban_correction_2024}, dark-field bias, interferometer vibrations \cite{noichl_correction_2024}, and beam hardening. Furthermore, the attenuation images were contrast enhanced and the dark-field images were denoised. Areas of direct x-ray exposure, surrounding the LUNGMAN, were concealed due to artifacts arising from detector non-linearity. The dark-field signal of the lung area, which was manually segmented based on the attenuation image, was averaged for the measured and simulated images.

\subsection{Simulated Dose Reduction Algorithm}
The dose reduction algorithm is a modification of the motion artifact correction algorithm originally developed for the clinical dark-field system to reduce or eliminate motion artifacts (e.g. from the heartbeat) \cite{Schick2022}. 
In summary, this algorithm  partitions the acquired data into multiple virtual scans, each simulating a reduced height of the grating area, which effectively shortens the illumination time and thereby mitigating motion-related artifacts. Through a chi-squared analysis the most suitable virtual scan that minimizes motion in each pixel is chosen for motion correction.

\begin{figure}[!ht]
    \centering
    \includegraphics[width=\textwidth]{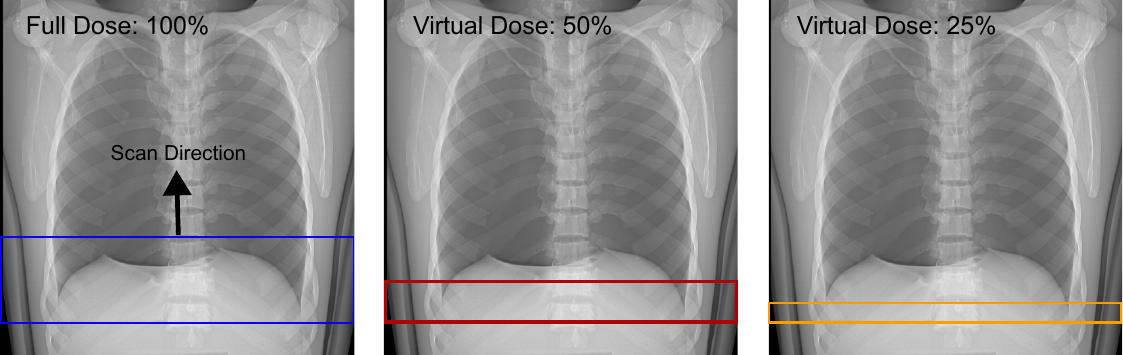}
   \caption[Simulated Dose Reduction Algorithm.]{Simulated Dose Reduction Algorithm. The dose reduction algorithm partitions the image data by virtually reducing the height of the original grating area (blue box) to \SI{50}{\%} (red box) or to 25\,\% (orange box), corresponding to dose reductions to 50\,\% or to 25\,\%, respectively.}
   \label{fig:algorithm}
   \vspace{2ex}
\end{figure}

However, this virtual reduction in grating height, as illustrated in Fig. \ref{fig:algorithm}, also corresponds to a simulated decrease in radiation exposure to both the detector and the patient, proportional to the extent of the height reduction. 
To enable simulated dose reduction, the algorithm was modified accordingly:
The height reduction of the grating area is no longer fixed but variable, allowing the simulation of different levels of dose reduction.
Additionally, instead of generating multiple virtual scans for the mitigation of motion artifacts, a single virtual scan is sufficient for the reconstruction of dark-field and attenuation radiographs with a reduced dose.

\subsection{Phantom Measurements}
To validate the dose reduction algorithm, measurements of the anthropomorphic chest phantom LUNGMAN (Kyoto Kagaku, Japan), filled with the accompanying urethane lungs, positioned the same way as patients, were taken. The two optional chest plates representing human fat tissue were attached to the phantom, thereby modeling a male person with a BMI of \SI{29}{\kg/\m^2}.
The phantom was imaged at \SI{456}{mA}, corresponding to the average radiation dose of a patient \cite{frank_dosimetry_2021a}, and further, reduced dose measurements at \SI{228}{mA} (50\,\%) and \SI{114}{mA} (25\,\%). 

\section{Results}
\begin{figure}[!ht]
    \centering
    \includegraphics[]{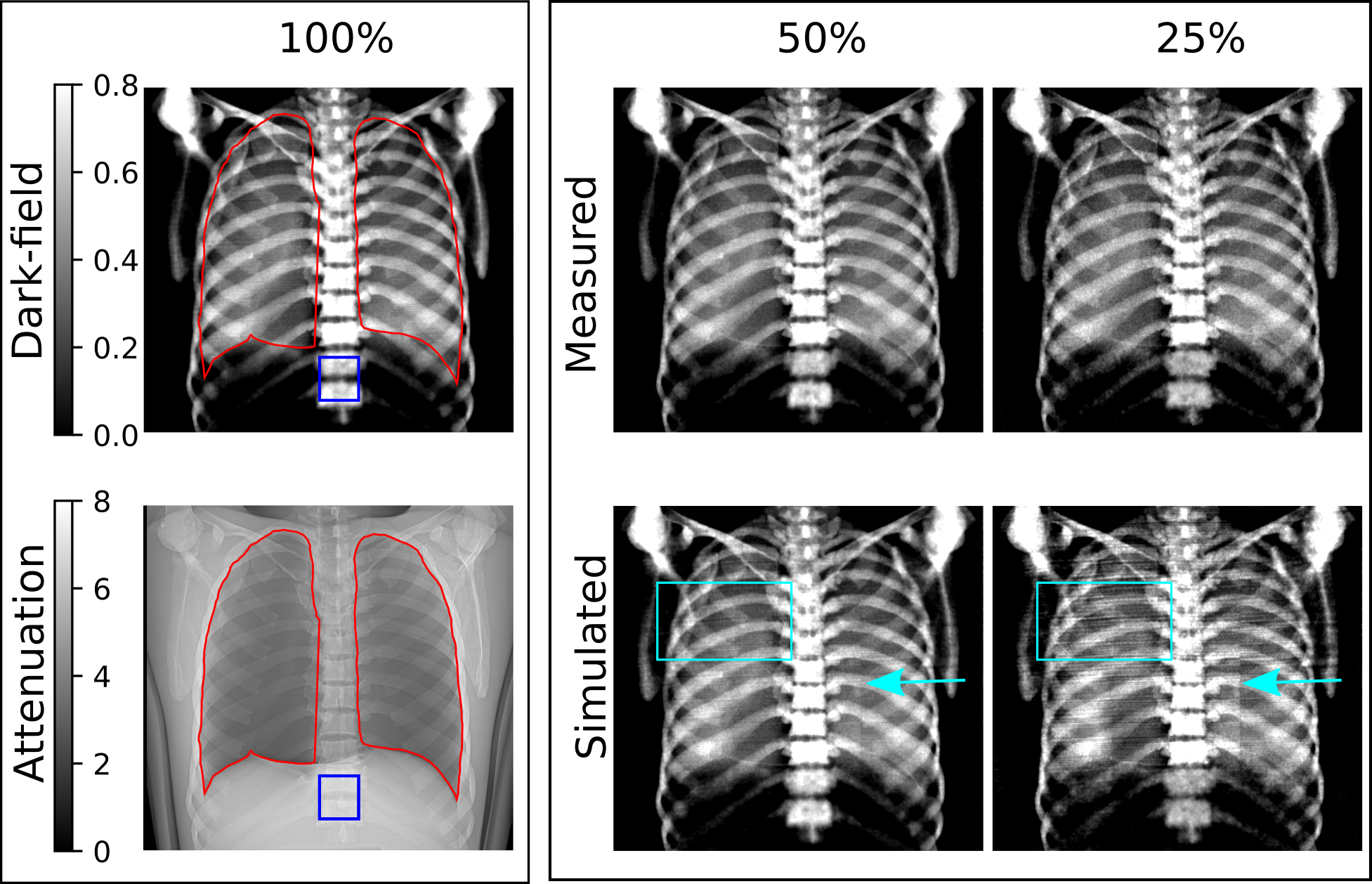}
\caption{The left part shows the dark-field and attenuation radiograph reconstructed from the patient-equivalent radiation dose. Red marks the lung area and the blue rectangle (dashed line) marks the area of the last two vertebrae. In both region-of-interest, the dark-field signal was averaged for further evaluation. The right part shows the measured and simulated dark-field images for half and quarter of the full dose. The turquoise rectangle (solid line) highlights an area with horizontal stripe artifacts and the turquoise arrow highlights a vertical line artifact at the stitching gap. \vspace*{2ex}} 
\label{fig:images}
\end{figure}

The patient-equivalent tube current of \SI{456}{mA} exposed the LUNGMAN phantom to an effective dose of \SI{87}{\micro Sv}. The reconstructed dark-field and attenuation image can be seen in the left part of \cref{fig:images}. 
As expected from a human lung phantom, the attenuation image closely resembles that of a real human. 
In contrast, however, the dark-field image does not fully replicate the dark-field image of a human. While the dark-field signal from the foam lungs is comparable to that of real lung tissue, the phantom's synthetic bones, which are composed of epoxy resin and calcium carbonate, generate a significantly stronger dark-field signal than actual human bones \cite{frank_dosimetry_2021a}, leading to a noticeable discrepancy.

Since the dark-field prototype system has a low sensitivity for osseous structures, the directional dark-field signal of bones can only made visible when the sample is positioned very close to the $G_1$ grating in order to maximize sensitivity \cite{gassert_darkfield_2023a}. 
For regular image acquisitions of patients, positioned in the patient's cabin, the sensitivity of the system is not high enough for measurements of the bone's dark-field signal.

The right part of \cref{fig:images} shows the measured and simulated dark-field radiographs for different dose levels. The general image appearance between measured and simulated radiographs is comparable: all images show the same antropomorphic LUNGMAN features with similar gray values (see table \ref{tab:df_values}).  The measured dark-field radiographs exhibit more noise with decreasing tube current, which can be seen as grainy texture in and in between the synthetic ribs. 
In contrast, the simulated dose-reduced dark-field images exhibit horizontal stripe artifacts that are not visible in the measured counterparts. These horizontal stripe artifacts are slightly visible in the \SI{50}{\%} dark-field image and worsen with further dose reduction. Looking closely at the measured full-dose image (\cref{fig:images} top left), one can also spot some horizontal stripes, which are not present in the measured low-dose images (\cref{fig:images} top right).
Furthermore, the simulated low-dose dark-field images show a vertical stripe artifact on the right-hand side, located at the stitching gaps of the individual G2 gratings (see \cite{frank_darkfield_2022}).

For a quantitative comparison between measurement and simulation, the average signal in the lung area and the signal of the last two vertebrae was analyzed (see \cref{tab:df_values}). In the measured images, the average dark-field signal of the masked lung area decreased by \SI{2}{\%} when reducing the tube current. The signal of the last two vertebrae decreased by up to \SI{15}{\%} when reducing the tube current.
The average dark-field signal in the simulated images partially mimics this decrease: the simulated images have a higher average dark-field signal in the lung area, which increased by \SI{1}{\%} when simulating \SI{25}{\%} instead of \SI{50}{\%} (see \cref{tab:df_values}). The simulated signal of the last two vertebrae drops by \SI{20}{\%} when simulating the \SI{22}{\micro Sv} image. The signal loss of the vertebrae can also be seen in \cref{fig:images}.
However, the quantitative differences between measurement and simulation cannot be seen visually when using the standard windowing. Also, the overall image appearance is dominated by the horizontal stripe artifacts making a quantitative comparison difficult (see \cref{fig:images}).

\vspace{1cm}

\begin{table}[!ht]
\centering
\begin{tabular}{lcccccccc}
             & & \multicolumn{3}{c}{Average Dark-field (Lungs)}                          &  \multicolumn{3}{c}{Average Dark-field (Vertebrae)} \\
 Current &  Effective dose & Measured             & Simulated             & Ratio  & Measured& Simulated &Ratio  \\
\toprule
    \SI{456}{mA}  &  \SI{87}{\micro Sv} & 0.389&     -     &  -            & 0.562 &-&-  \\
  \SI{228}{mA}   &\SI{44}{\micro Sv}& 0.382  &  0.402      &  1.05    & 0.513 &0.566 &1.10\\
      \SI{114}{mA} & \SI{22}{\micro Sv} & 0.383 &   0.406 & 1.06 & 0.475 &0.451 &0.95\\
\bottomrule
\end{tabular}
\label{tab:df_values}
\caption{Effective dose for the different tube currents and the average dark-field signal in the region of the lungs and the two last vertebrae for measured and simulated radiographs together with the ratio of both quantities.\vspace{4ex}}
\end{table}

\section{Discussion}
In this work, we compared the radiographs of a chest phantom generated by simulated dose reduction using the algorithm proposed by Schick \& Bast et~al. \cite{schick_simulated_2024} and by actual dose reduction through variations of the X-ray tube current.
The algorithm proposed by Schick \& Bast et~al. \cite{schick_simulated_2024} simulates a reduced  radiation dose by virtually reducing the illuminated area, thus reducing the amount of radiation that would have reached the patient or any sample. 

A real reduction of the radiation dose at the prototype system for clinical dark-field radiography can be achieved in different ways:
(1) A variation of the X-ray tube current; 
(2) a reduction of the X-ray beam collimation during the image acquisition scan, equivalently to the simulated dose reduction, and 
(3) changes of the image acquisition procedure (e.g, scanning speed, length or number of the X-ray pulses), which cannot be implemented without significant efforts.
The first method was chosen for the measurement of reduced dose images of the phantom due to both its simplicity and absence of significant drawbacks.
An algorithm, employing this method for dose reduction as-well is feasible, but requires the simulation and modeling of accurate noise characteristics.

The second method, which was employed for our dose-reduction algorithm, has the benefit of simple software implementation, as no noise modeling is necessary. Furthermore, this method conserves the noise characteristics of scans with high radiation dose, as the impact of electronic noise per image frame is low compared to scans with low tube currents.
However, for higher levels of dose reduction, reducing the irradiation area also decreases the number of stepping positions for the image reconstruction, thus leading to a worse phase coverage of the stepping curve and, therefore, eventually to horizontal stripe artifacts.  
These algorithm-associated horizontal stripe artifacts, which arise at simulated dose reductions at around 50\,\% are not visible in their measured counterparts (cf. Fig. \ref{fig:images})

Quantitatively, the average dark-field signal differs by \SI{5}{\%} to \SI{10}{\%} from the measured images. However, these differences are not observable for the clinical dark-field images with a regularly employed windowing of 0 to 0.8. 
The observed horizontal stripe artifacts, however, have a greater visual impact and are disrupting the overall image appearance more severely than a slight change in the mean dark-field intensity.  
Since the measured low-dose images do not show the same stripe artifacts as the simulated radiographs, the image quality of such low-dose radiographs is subjectively better than their simulated counterparts.

\section{Conclusion}
Our results demonstrate that dark-field images obtained at \SI{22}{\micro Sv}, which corresponds to a quarter of the current target dose for persons with a BMI of \SI{29}{\kg/\m^2}, can be reconstructed artifact-free. When simulating low-dose radiographs by virtually reducing the irradiated area, arising stripe artifacts are purely algorithm-induced. 

These results confirm the potential of an even higher possible dose reduction without algorithm-associated artifacts as indicated in our previous study \cite{schick_simulated_2024} for the diagnosis of COVID-19 without loss of diagnostic accuracy.

\section{Funding}
We acknowledge financial support through the European Research Council (ERC Synergy Grant SmartX, SyG 101167328), and the Free State of Bavaria under the Excellence Strategy of the Federal Government and the Länder, as well as by the Technical University of Munich – Institute for Advanced Study.

\section{Competing Interests Statement}
T.K. is an employee of Philips Innovative Technologies in Hamburg, Germany. The other authors have no conflict to disclose.

\printbibliography

\end{document}